\documentclass{ws-p8-50x6-00}
\def\epem{e^+e^-}

\def\as{\alpha_{\mbox{\scriptsize s}}}

\def\de{\delta}

\def\out{\mbox{\scriptsize out}}

\def\cO#1{{\cal{O}}\left(#1\right)}

\def\abs#1{\left| \: #1 \: \right|}

\def\PT{\mbox{\scriptsize PT}}

\def\conf{\delta}

\def\Ko{K_{\out}}

\def\cD{{\cal{D}}}

\def\cM{{\cal{M}}}
\def\cS{{\cal{S}}}

\def\cC{{\cal{C}}}

\def\sl{r_{\SL}}

\def\cV{{\cal {V}}}

\def\cD{{\cal {D}}}

\def\Journal#1#2#3#4{{#1} {\bf #2}, #3 (#4)}
\def\NPB{{\em Nucl. Phys.} B}
\def\PLB{{\em Phys. Lett.}  B}

\def\JHEP{\em JHEP}

\def\epj{{\em Eur. Phys. J.} C}

\def\PREP{\em Phys. Rep.}

 \newskip\humongous \humongous=0pt plus 1000pt minus 1000pt
   \newif\ifdtup

\def\fun#1#2{\lower3.6pt\vbox{\baselineskip0pt\lineskip.9pt
  \ialign{$\mathsurround=0pt#1\hfil##\hfil$\crcr#2\crcr\sim\crcr}}}
\begin{document}

\title{Interplay between perturbative and non-perturbative QCD in
  three-jet events}

\author{Andrea Banfi}

\address{Dipartimento di Fisica G. Occhialini, Universit\`a di
  Milano-Bicocca,\\ P.za della Scienza 3, 20126 Milano, Italy\\
  E-mail: andrea.banfi@mib.infn.it}

\author{Giulia Zanderighi}

\address{Dipartimento di Fisica Nucleare e
  Teorica, Universit\`a di Pavia,\\ Via U. Bassi 6, 27100 Pavia, Italy\\
  E-mail: zanderighi@pv.infn.it}


\maketitle

\abstracts{ We present the perturbative (PT) and non-perturbative (NP)
  analysis of the cumulative out-of-event-plane momentum distribution
  in $\epem$ annihilation in the near-to-planar three-jet region.
  A physical interpretation based on simple QCD considerations and
  kinematical relations will be given, with the aim of extending the
  described techniques to other multi-jet processes and, possibly, to
  hadron-hadron collisions.}

\section{Introduction}
Hadronic multi-jet events play a crucial r\^ole both in the context of
precision tests of QCD and in the search for new physics, so that it
has become essential to reach, in the analysis of multi-jet
configurations, the same theoretical accuracy as in two-jet events.
As a first step in this direction we aim to extend the
``state-of-the-art '' of two-jet event shape variables such as Thrust
(T), C-parameter, Broadening (B), to three-jet event shapes.  This
standard analysis consists in a resummed single logarithmic (SL)
prediction, the exact fixed order result, the matching of the two and
the non-perturbative power corrections.

We present here the (Thrust) Minor distribution, which gives a measure
of the aplanarity of a three-jet event.  The final answer is rather
involved, since it reveals the rich colour and geometry structure of
the hard underlying process ($e^+ e^- \rightarrow q\,\bar{q}\, g$).

The aim of this paper is to present only the main features of the
Minor distribution, while all computational details may be found in a
separate paper.\cite{aco} In section 2 we introduce the observable and
the distribution.  The physical interpretation of the SL resummed PT
result is explained in section 3. Section 4 is devoted to the power
corrections.  We conclude in section 5 giving some outlooks.

\section{Observable and kinematics}
The (Thrust) Minor ($T_m$) gives a measure of the cumulative
out-of-event-plane momentum $\Ko$:
  \begin{equation}
  \label{eq:thrustm}
\hspace{-0.8cm}
  T_m\,Q = \sum_{i} \abs{{p}_{ix}} \>\equiv\> \Ko\>.
\end{equation}
Here Q is the centre-of-mass energy, the sum is over all hadrons and
we have fixed the z- and the y-axis along the thrust ($T$) and the
thrust-major ($T_M$) axes respectively\cite{asmeasure}.

At Born level a three-jet event consists of a quark, an antiquark and
a hard, non-collinear gluon.  For kinematical reasons these partons
lie in a plane, so that $\Ko=0$.  We denote by $p_a$ the energy
ordered ($p_1^0 > p_2^0 > p_3^0$) parton momenta. There are
essentially three Born configurations: we denote by $\conf$
($\conf=1,2,3$) the configuration in which the momentum of the hard
gluon is $p_{\conf}$ (see Fig.~1).
\begin{figure}[ht]
  \begin{center}
    \epsfig{file=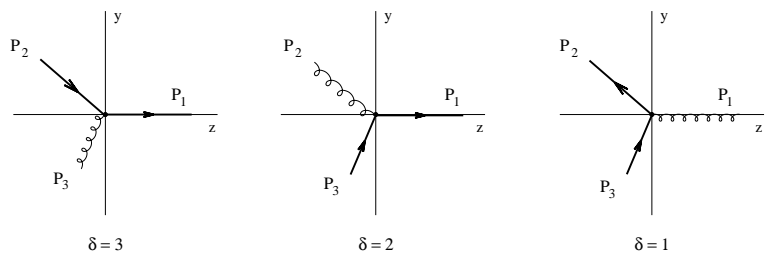,width=0.85\textwidth} 
    \label{fig:conf}
    \caption{The three Born configurations ordered according to
      decreasing probability.}
  \end{center}
\end{figure}

Beyond Born level one can study the ``integrated'' $\Ko$-distribution
$\Sigma_{\conf}(\Ko)$, defined by
\begin{equation}
  \label{eq:DLdist}
\hspace{-0.4cm}
\frac{d\sigma_{\conf}(T,T_M,\Ko)}{dT\>dT_M\>d\Ko}
\equiv\>M_{\conf}^2(T,T_M)\frac{d \cV_{\conf}(\Ko)}{d\Ko}\>,
\end{equation}
\begin{equation}
  \label{eq:DLdist2}
\hspace{-0.4cm}
\cV_{\conf}(\Ko) \equiv \cC(\as) \Sigma_{\conf}(\Ko)+\cD_{\conf}(\Ko)\>.
\end{equation}
Here $M_{\conf}^2(T,T_M)$ denotes the hard matrix element for the
production of a quark-antiquark-gluon ensemble in the configuration
$\conf$ for fixed $T$ and $T_M$.  $\Sigma_{\conf}(\Ko)$ resums all
double (DL) and all single logarithms (SL). Hard emitted parton
contributions\cite{CTTW} are embodied both in the ``coefficient
function'' $\cC(\as)$, which has an expansion in powers of $\as(Q)$,
and in the ``remainder function'' $\cD(\Ko)$, which vanishes for
$\Ko\rightarrow 0$.
\section{PT result}
The PT resummed result to SL accuracy
has the following structure\cite{aco}
\begin{equation}
  \label{eq:PT-res}
\Sigma_{\conf}(\Ko) = e^{-R(\Ko)}\cdot \cS_{\conf}(\Ko)\>.   
\end{equation}
Here $R$ is a DL function 
which resums all soft and collinear parton emissions,  
\begin{equation}
  \label{eq:rad-dl}
  R(\Ko)=\sum_a C_a
  \int_{\Ko}^{Q_a^{\PT}}\frac{dk_x}{k_x}
\frac{2 \as(2k_x)}{\pi}\ln\frac{Q_a^{\PT}}{k_x}
\stackrel{DL}{\longrightarrow}
C_T\frac{\as(Q)}{\pi}\ln^2\frac{Q}{\Ko}\>,
\end{equation}
where $C_a$ is the colour charge of parton $\# a$ ($C_{\conf}=C_A, C_a
= C_F $ for $ a \neq \conf$) and $C_T=2C_F+C_A$ is the {\em total}
colour charge of the hard quark-antiquark-gluon ensemble.  Sources of
SL corrections in $R$ are the running of the coupling, corrections due
to hard collinear splittings and the dependence on the geometry
through the three hard momentum scales $Q_a^{\PT}$.  Each of these
scales has a nice geometrical interpretation: for the quark or the
antiquark it is (proportional to) the $q\bar{q}$ invariant mass, for
the hard gluon it is (proportional to) its invariant transverse
momentum with respect to the $q\bar{q}$ dipole.

\begin{figure}[ht]
  \begin{center}
    \epsfig{file=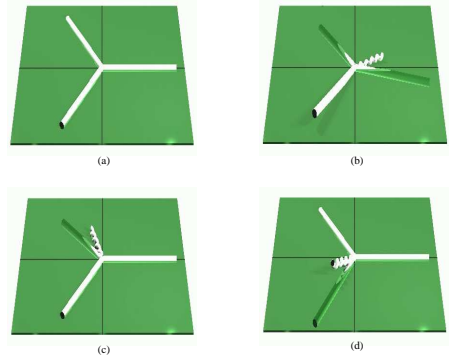,width=0.52\textwidth}
   \caption{
     (a) Hard partons in a generic Born
     configuration.
     (b) Soft gluon $k$ (the short curly
     stick) is emitted from $p_1$. All three hard partons
     experience equal out-of-plane recoils, see Eq.\ref{2.8}. White
     (shadowed) partons have positive (negative) $x$-components of the
     momentum.
     (c) Soft gluon $k$ is emitted from $p_2$.  According to
     Eq.\ref{2.9}, only parton \#2 recoils (shadowed; has a negative
     $x$-component of the momentum), parton \#1 and \#3 (white) remain
     in the event plane.  (d) The case of $k$ emitted from $p_3$.
\label{fig:recoil}}
  \end{center}
\end{figure}

The remaining SL corrections, in particular those due to hard parton
recoil, are embodied in the SL function $\cS$.  At first order in
$\as$ (one secondary gluon emission), $\cS$ is given by
\begin{equation}
  \label{eq:cs}
  \cS=1-\frac{2\as}{\pi}\ln \frac{Q}{\Ko}(2 C_1 + C_2 +C_3)\ln 2\>.
\end{equation}
This shows that the contribution to $\Ko$ due to emission from
the parton along the thrust axis ($p_1$) is twice the contribution of
each of the remaining two emitters.  There is a simple kinematical
reason for this.\cite{aco} Indeed the definition of the $T$- and
$T_M$-axes implies that when the secondary gluon $k$ is emitted from
$p_1$ (see Fig.~\ref{fig:recoil}b) all three hard partons experience
equal out-of-plane recoils,
\begin{eqnarray}
\label{2.8}
k_z>0, \quad & k_y>0: \quad   p_{1x}=-k_x=p_{2x}=-p_{3x}\quad\\
       & k_y<0: \quad   p_{1x}=-k_x=-p_{2x}=p_{3x}\>,\nonumber
\end{eqnarray}
so that $\Ko=4\cdot \abs{k_x}$.
On the other hand, when the secondary gluon is emitted 
from $p_2$ or $p_3$ (see Fig.~\ref{fig:recoil}c or d) 
only one hard parton recoils against it:
\begin{eqnarray}
\label{2.9}
k_z<0, \quad &  k_y>0: \quad   p_{2x}=-k_x; \>\>  p_{1x}=p_{3x}=0\quad \\
             &  k_y<0: \quad   p_{3x}=-k_x; \>\> p_{1x}=p_{2x}=0\>,\nonumber
\end{eqnarray}
so that $\Ko=2\cdot |k_x|$.

\section{NP result}
Power corrections arise because of emission of extra-soft gluons
with transverse momentum of order $\Lambda_{QCD}$. In this phase space
region a pure PT approach is not possible. 
In fact, due to the growth of the PT coupling at low scales, the PT
series have a factorial behaviour, see a recent review for details.
\cite{renormalons}
These effects (both PT and NP) are needed for
precision tests of QCD and are a way to explore low energy regions.

The {\em Ansatz} we start from is that there exists an IR-finite
running coupling defined at all scales in terms of a dispersive
representation \`a la DMW.\cite{DMW} For a generic $\epem$
event shape variable ($V$) power corrections result in a shift of the
PT distribution
\begin{equation}
\Sigma(V)=\Sigma^{\PT}(V-\delta V)\>.
\end{equation}
The NP part of the shift has the following general
form\cite{thr-milan}:
\begin{equation}
\delta V=
\rho(V) \cM\frac{\mu_I\alpha_0}{Q}+\cO{\as(Q)\frac{\mu_I}{Q}}\qquad 
\alpha_0(\mu_I)= \frac{1}{\mu_I}
\int_0^{\mu_I}\! dk\, \alpha_s(k)\>.
\end{equation}
Here $\alpha_0$ is the universal parameter which measures the
strength of the strong interaction at low scales: it is the mean value
of the full coupling constant (PT and NP) below a certain merging
scale.  $\cM$ is the Milan factor which takes into account two-loop
corrections in a universal way.  The coefficient function $\rho(V)$
is specific of the observable considered and can be computed via
PT calculations.

As in the case of Broadening\cite{broad}, the shift of the Minor
distribution depends logarithmically on the hard parton recoil momenta
($p_{ax}$)
\begin{equation}
\label{eq:deltako}
\Delta\Ko\sim \sum_a C_a\ln\frac{Q}{|p_{ax}|}\>,
\end{equation}
so that, once the shift $\Delta(\Ko)$ in a generic PT
configuration is known, the total shift is given by the average
of its expression over the PT recoil distribution:
\begin{equation}
\delta\Ko\,= \, \langle \Delta\Ko\rangle\>_{\PT}\>. 
\end{equation}
There are two interesting limiting cases, the region of well developed
(multiple) PT radiation, i.e. $\as\ln^2(\Ko/Q)\gg 1$ and the phase space
region with only few hard PT partons, i.e. $\as\ln^2(\Ko/Q)\ll 1$. 

In the first case all three hard partons have similar recoils
($p_{ax}\sim \Ko$) and the shift results in
\begin{equation}
\de\Ko\sim C_T \ln \frac{ Q}{\Ko}\>.
\end{equation}

In the second case the shift depends on which hard parton controls the
PT emission and therefore the recoil. It is therefore useful to
single out in the total shift the contribution of each hard parton:
\begin{equation}
  \label{eq:shift}
  \delta\Ko=\sum_a \frac{C_a}{C_T}\delta\Ko^{(a)}\>.
\end{equation}
Here $C_a/C_T$ represents (roughly) the probability that PT radiation is due to
emission from parton $ \# a$. 

\begin{itemize}
\item PT emission from $p_1$: all three hard partons have similar
  out-of-plane recoils (see Eq.\ref{2.8}), so that, as in the previous
  case, $p_{ax}\sim \Ko$ and the shift is a logarithm
\begin{equation}
\delta \Ko^{(1)} \sim (2C_F+C_A)\ln\frac{Q}{\Ko}\>.
\end{equation}
\item PT emission from $p_2$: only the out-of-plane momentum $p_{2x}$
  is fixed by PT radiation (see Eq.\ref{2.9}), so that $p_{2x}\sim
  \Ko$ and one has to integrate over the PT distribution of the other
  two ``free'' hard partons:
\begin{equation}
\delta \Ko^{(2)} \sim \frac{C_F}{\sqrt{\as C_F}}+C_F\ln\frac{Q}{\Ko}
+\frac{C_A}{\sqrt{\as (C_F+C_A)}}\>.
\end{equation}
In this case the shift contains a $1/\sqrt{\as}$ enhancement coming
from the average of $\ln Q / p_{ax}$ ($a=1,3$) in Eq. \ref{eq:deltako}
over the corresponding DL Sudakov form factor.\cite{aco} Similar
considerations hold for PT emission from $p_3$.
\end{itemize}
\section{Conclusions and outlook}
Special features of Thrust Minor distributions are the richness of the 
geometry dependent structure and the sensitivity to large angle soft
gluon emissions. 

The next step will be to extend this calculation to hadron-hadron
collisions, where one has to take into account effects due to initial
state radiation.

\section*{Acknowledgements}
The work here presented has been done in collaboration with Pino
Marchesini and Yuri Dokshitzer.  We are thankful to Gavin Salam for
helpful discussions and suggestions.  One of us (G.Z.) would like to
express her gratitude towards the organisers of the ISMD conference
for the stimulating and pleasant atmosphere.

\end{document}